\documentclass[12pt,preprint]{aastex}

\shorttitle{Kinematics of PNe with low mass progenitors}
\shortauthors{Pereyra et al.}

\begin{document}


\title{The kinematics of the nebular shells around low mass progenitors of PNe with low metallicity \footnote{The observations reported herein were acquired at the Observatorio Astron\'omico Nacional in the Sierra San Pedro M\'artir (OAN-SPM), B. C., Mexico.}}


\author{Margarita Pereyra\altaffilmark{1,2}, Jos\'e Alberto L\'opez\altaffilmark{1} \& Michael G. Richer\altaffilmark{1} }
\affil{ \altaffilmark{1}Instituto de Astronom\'\i a, Universidad Nacional Aut\'onoma de M\'exico, \\ Apartado Postal 106, C.P. 22800 Ensenada, BC, M\'exico} 
\affil{ \altaffilmark{2}Schlumberger Foundation Fellow, School of Physics and Astronomy, \\ University of Southampton,  Southampton, SO17 1BJ, UK }
\email{mally@astrosen.unam.mx, jal@astrosen.unam.mx \&  richer@astrosen.unam.mx }



\begin{abstract}

We analyze the internal kinematics of 26 Planetary Nebulae (PNe) with low metallicity that appear to derive from progenitor stars of the lowest masses, including the halo PN population. Based upon spatially-resolved, long-slit, echelle spectroscopy drawn from the San Pedro M\'{a}rtir Kinematic Catalogue of PNe \citep{Lopez12}, we characterize the kinematics of these PNe measuring their global expansion velocities based upon the largest sample used to date for this purpose.  We find kinematics that follow the trends observed and predicted in other studies, but also find that most of the PNe studied here tend to have expansion velocities less than 20\,km\,s$^{-1}$ in all of the emission lines considered.  The low expansion velocities that we observe in this sample of low metallicity planetary nebulae with low mass progenitors are most likely a consequence of a weak central star wind driving the kinematics of the nebular shell.  This study complements previous results \citep[and references therein]{Pereyra13} that link the expansion velocities of the PN shells with the characteristics of the central star.  
\end{abstract}





\section{INTRODUCTION}\label{sec_introduction}

 Our current theoretical understanding of the kinematics of the nebular shells of PNe clearly predicts an evolution of the kinematics with time \citep{Kwok82, Mellema94, Villaver02, Perinotto04, GS06, Schonberner10}. According to these models, the kinematic evolution of the nebular shells in PNe depends basically on two factors: the mass loss on the AGB and the energy provided by the central star (CS) through its wind and ionizing radiation field.  Recently, \citet{Richer08, Richer10} and \citet{Pereyra13} have provided firm observational support for these predictions. Using large samples of PNe segregated by evolutionary stages, they presented for the first time the complete evolutionary sequence for the global kinematics of the nebular shell over the lifetime of the PN as a function of the CS evolution. The global expansion or bulk outflow velocity can be adequately determined from the matter with the highest emission measure within the spectrograph slit, in contrast with the theoretical expansion velocity sometimes referred as pertaining to the shell's outer post-shock velocity, which cannot be measured directly \citep{Schonberner05b}. The global expansion velocity (referred simply as expansion velocity hereafter) of the nebular shell steadily increases until the CS reaches its maximum effective temperature and ceases nuclear burning. Thereafter, the expansion velocity of the shell is observed to slow down as the luminosity and wind power of the CS rapidly falls and the CS approaches the white dwarf cooling track.  At this advanced stage of evolution there may be additional effects contributing  to the observed deceleration of the nebular shell, such as ionization stratification, decreasing optical thickness, interaction with the surrounding medium and, possibly, recombination of the outer shell regions in some cases \citep[e.g.][]{Wareing07, Sabin10, Schonberner14, Pereyra13}. 

Here, we extend our previous studies to PNe derived from stellar progenitors of lower masses than those of the typical PNe in the disc or bulge of the Milky Way. To this end we draw the largest sample used to date for this purpose from the San Pedro M\'{a}rtir Kinematic Catalogue of PNe \citep[henceforth, SPM Catalogue;][]{Lopez12} to perform the kinematic analysis. Our aim is to characterize the role played by the CS in driving the expansion velocity of the nebular shells for the progenitor stars of the lowest masses, seeking in general CS progenitor masses $\leq$ 1 M$_{\odot}$, which are also the oldest and of low metallicity as a result of the chemical evolution of the Milky Way \citep[e.g.,][]{bensbyetal2013, bensbyetal2014}. The sample contains all of the PNe (11 objects) currently considered as belonging to the Galactic halo plus a group of very low metallicity PNe (log(O/H) + 12 $\lesssim 8.0$ dex), for a total of 26 objects. Although the size of the sample is modest this should provide a fair representation of the typical kinematic behavior of the target group under study.

The selection criteria, data, and the methods of measurement are described in \S 2. In \S 3, we present the results for the kinematic behavior of the halo and the low metallicity groups and the analysis of the velocities obtained for the entire sample. Finally, in \S \ref{sec_discussion} and \S \ref{sec_conclusions} we discuss the results and summarize our conclusions. 

\section{THE SAMPLE}

\subsection{Selection Criteria and the Data}

All of the kinematic data considered here are drawn from the SPM Catalogue \citep{Lopez12}\footnote{http://kincatpn.astrosen.unam.mx}, which contains high resolution spectra for nearly 700 Galactic PNe observed from the Observatorio Astron\'omico Nacional in the Sierra de San Pedro M\'artir (OAN-SPM) in Ensenada, B.C., M\'exico.  The declination range accessible from SPM spans from $+70 ^\mathrm{o}$ to $-40 ^\mathrm{o}$. The observations were obtained with the Manchester Echelle Spectrograph (MES). This is a long slit echelle spectrograph without cross-dispersion \citep{meaburnetal1984,meaburnetal2003}. Narrow-band filters isolate orders 87 and 113 covering approximately the wavelength ranges 6540 - 6595 \AA ~and 4990 - 5030 \AA, respectively, that contain the  H$\alpha$, [\ion{N}{2}] $\lambda$$\lambda$6548,6584, and [\ion{O}{3}] $\lambda$ 5007 emission lines.  The slit is $\sim6\farcm5$ long and, for all of the observations considered here, a width of 150 $\mu$m ($\sim 1\farcs9$) was used, resulting in a spectral resolution of 11\,km\,s$^{-1}$ FWHM, easily resolving the internal kinematics of typical PNe. The spectra are calibrated using exposures of a ThAr lamp, with an accuracy of $\pm1$\,km\,s$^{-1}$ when converted to radial velocity.  \cite{Lopez12} provide futher details regarding the data reduction.  Most of the objects in the present sample were observed at a single slit position passing close to the center of the nebula.  For the few objects observed in several positions, we used the slit position passing closest to the center of the nebula.  

The known Galactic population of PNe amounts to about 3500 objects and is largely concentrated in the Galactic plane \citep{Parker06}. Most of these PNe have central star masses  $\geq 1$ M$_{\odot}$ whereas those PNe of small core mass spread out of the plane and into the inner regions of the Galactic bulge \citep[e.g.,][]{CorradiSchwarz95, Zhang&Kwok97, Cavichia11}.  The age - metallicity relation for central stars of PNe has been discussed by  e.g., \citet{Maciel10}. The results in these and other works of Galactic chemical evolution indicate that the lower the central star mass, the lower the metallicity and the slower the stellar evolution.  

In order to obtain a reliable sample of PNe derived from low-mass progenitors ($\leq 1$ M$_{\odot}$), we first selected all the PNe that have previously been classified as belonging to the Galactic halo, including those in globular clusters, and whose CSs are considered descendants of Pop II stars.  This sample of PNe corresponds to type IV PNe \citep{Peimbert90} and are characterized by their low metallicity relative to disc PNe and are associated with slowly evolving, low-mass  stars.  The halo is not the only Galactic component with low mass stars.  The Milky Way contains old stars that are also metal-rich \citep[e.g.,][]{bensbyetal2013}, particularly in the Bulge.  Unfortunately, we have no way of discriminating their PN progeny from PNe descended from younger, metal-rich stars of higher mass, so it is not clear how we might include them without contaminating the sample.  Selected in this way, our halo sample includes 11 objects currently considered as PNe from the Galactic halo.  

Since the halo sample will be biased to PNe derived from progenitor stars of low mass and low metallicity, we chose to increase the sample following the same premise to select PNe from other Galactic stellar populations.  We compiled a list of PNe with low oxygen abundances from the literature, which should also descend from old, low-mass progenitors.  There is a tension between the choice of the upper limit to the chemical enrichment for inclusion in this study and the need to obtain a sample of a reasonable size.  Considering the $\mathrm O/\mathrm H$ abundance ratio for the Sun $\log(\mathrm O/\mathrm H) + 12 =8.7$\,dex; \citep{MA08}, as well as for PNe in the bulge and disc of the Milky Way 
$\log(\mathrm O/\mathrm H) + 12 =8.6$\,dex;  \citep{STS98,Kwok00}, and the average oxygen abundances for the halo PNe (see Table \ref{table02}), $\log(\mathrm O/\mathrm H) + 12 =7.8$\,dex, we consider low metallicity PNe as those whose oxygen abundance is $\log(\mathrm O/\mathrm H) + 12 \lesssim 8.0$\,dex, i.e., an oxygen enrichment five times lower than in the Sun.  There are considerable uncertainties and differences in the methods used by different authors to determine chemical abundances that hamper compiling a reliable or consistent sample from different sources. However, our limiting oxygen abundance allows uncertainties of up to 0.3 dex in abundance determinations and still yields a sample of PNe with low metallicity relative to the Galactic disc and bulge.  We only consider oxygen abundances derived from collisionally-excited emission lines.  Presently, there is no broadly-accepted explanation for the discrepancy between the chemical abundances measured from recombination and collisionally-excited lines \citep[e.g.,][]{peimbertpeimbert2006, liu2012}.  However, there exist more measurements based upon collisionally-excited lines, so these were adopted in the hope of enlarging our sample.




During the selection process, we noted that some low metallicity PNe presented very wide H$\alpha$ line profiles, sometimes with a prominent red continuum, very similar to those observed in the spectra of D-Type symbiotic stars or young nebulae surrounding Post-AGB stars with dense outer H I envelopes.  Since D-Type Symbiotic stars are interacting binaries consisting of a hot compact star and a cool giant with a dusty envelope, these objects cannot be considered as true PNe. In these objects the giant component loses mass, part of which is  ionized by its hot companion, giving rise to nebular emission resembling a PN. However, the outer neutral hydrogen component scatters Ly$\beta$ photons that emerge as wide wings in the H$\alpha$ profile (known as Raman scattering) that are not observed in the spectra of PNe \citep[for further details, see e.g.,][]{Lee00, Corradi95}.  A similar situation arises in young Post-AGB nebulae given the amount of neutral hydrogen that can be present in the outer envelope  of these objects at this stage.  Although they are very different objects, their manifestation is similar in our high resolution spectra, via very wide H$\alpha$ lines.  According to our data, the symbiotic objects are H 1-45, H 2-43, Hen 2-171, and M 2-24, while the very young PNe or post-AGB objects are M 1-37, M 2-29, and M 2-39.  These objects deserve further scrutiny, which is beyond the scope of the present paper, but are not considered in what follows, as they were not included in our previous studies \citep{Richer08, Richer10, Pereyra13}.  

We further noticed some peculiar spectral line profiles indicative of bipolar outflows in the planetary nebulae K 4-47 and M 3-2. 
No other object in the sample shows line profiles with indications of bipolarity.  However, since many of the objects in the sample are spatially-unresolved under seeing-limited conditions from the ground, we decided to check their morphology searching for high resolution images available in the Hubble Planetary Nebula Image Catalogue compiled by Bruce Balick (http://www.astro.washington.edu/users/balick/PNIC/) and the Hubble Legacy Archive. We found images in these repositories for 17 objects of the sample. In these images, we observe indications of bipolar structure present, in addition to K 4-47 and M 3-2, in three other objects, K 5-17, M 1-6 and NGC 6644.  Their bipolar nature was not detected from their spectra because of their compact sizes and the orientation of the spectrograph slit. 
Given that bipolar PNe are usually associated with higher mass progenitors \citep[e.g.,][]{Peimbert90} and so not likely the descendents of old, low mass stars, we decided to exclude these objects from our sample.  Our final sample of low metallicity PNe includes 15 objects in addition to the 11 objects from the Halo.  

We recall that our purpose is to characterize the general evolution of PNe derived from stars of low mass and practical considerations impose the need to select for low metallicity as well, which clearly limits our sample to a more extreme sample than if we could select it only for low mass progenitor stars.  Even so, it is a valid sample that explores a region of parameter space that has been little explored previously.  It is likely that examples of PNe derived from low mass, but more metal-rich progenitor stars are already included in the studies of \citet{Richer08, Richer10}.

Table \ref{table02} presents our final sample of 26 objects that includes the halo PNe and apparently bona-fide, low metallicity PNe with kinematic data available from the SPM Catalogue \citep{Lopez12}.
The first two columns present the common and PNG name for each object.  Columns $3$ through $5$ present our measurements of the expansion velocity ($\mathrm V_{exp}$), from H$\alpha$, [\ion{N}{2}]\,$\lambda$6584, and [\ion{O}{3}]\,$\lambda$5007. Columns $6-8$ indicate the [O/H] abundance values, their references and whether they belong to the \lq\lq halo" or the \lq\lq low metal" samples. In columns $9$ and $10$, we list the \ion{He}{2}\,$\lambda$4686 intensities for each PN and their references, while columns $11-13$ show the luminosity and effective temperature of the CS with their references. 

Although it is not obvious that it has any consequence in our analysis that follows, we note that half of the PNe in Table \ref{table02} have a measured intensity for the \ion{He}{2} $\lambda$4686 line.  





\subsection{Measurements}

The calibrated long-slit spectra were analyzed using the IRAF\footnote[1]{{IRAF is distributed by the National Optical Astronomy Observatories, which are operated by the Association of Universities for Research in Astronomy, Inc., under cooperative agreement with the National Science Foundation.}} package to obtain the expansion velocity for the nebular shell.  
For a spatially-resolved PN undergoing spherical expansion, the kinematics are represented by a velocity ellipse in a bi-dimensional spectrum or position - velocity ({\it P--V}) diagram.  
Given the nature of our sample, many of the PNe are located far from the solar neighborhood (e.g., the Bulge) and thus they are not spatially resolved from the ground by the spectrograph slit. Depending upon the emission lines detected for each PN in our sample, we obtained velocity measurements from [\ion{N}{2}]\,$\lambda$6584, [\ion{O}{3}]\,$\lambda$5007, and H$\alpha$\,$\lambda$6563. We derive the expansion velocity of the nebular shell from the main receding (redshifted) and approaching ( blue-shifted) components of the line profile from the matter with the highest emission measure.   

\begin{figure}[]
\begin{center}
%
%
\includegraphics [width=0.12\columnwidth]{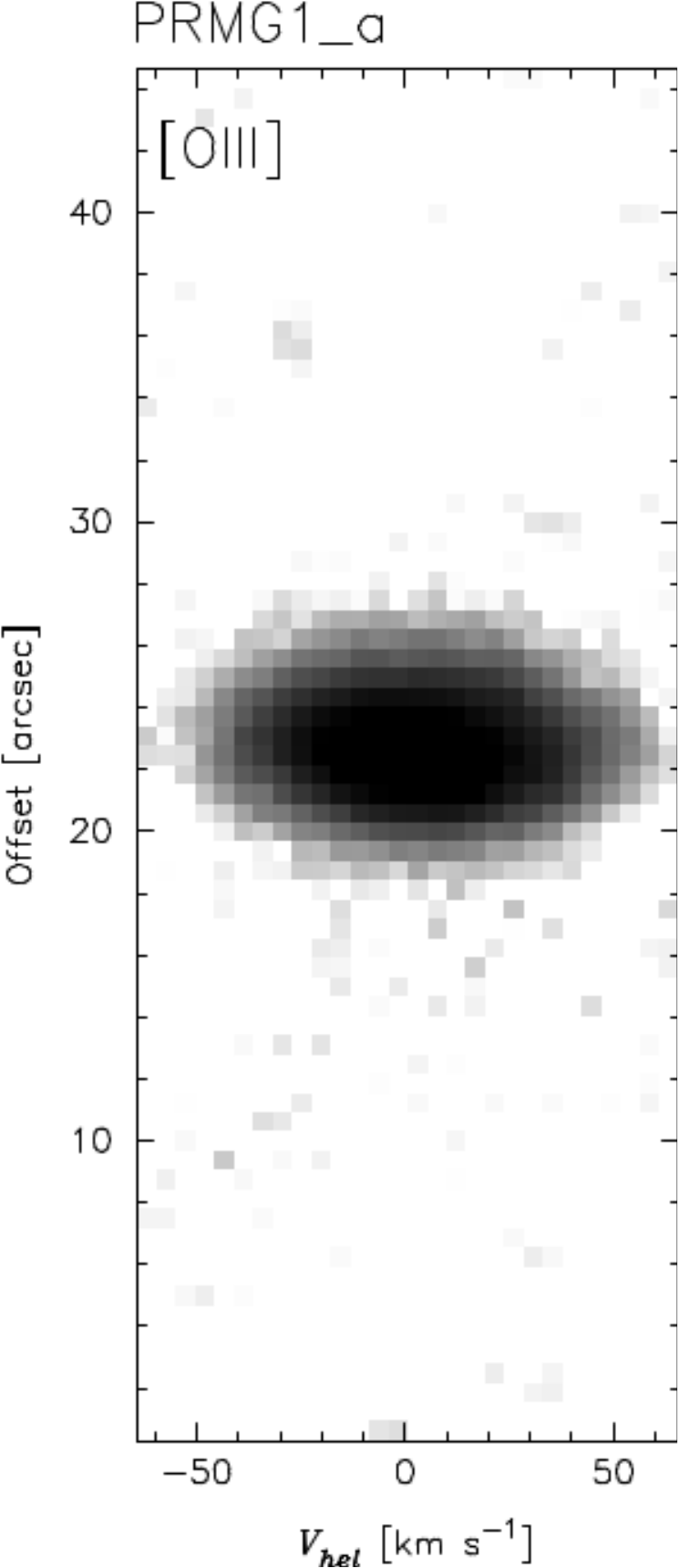} 
\includegraphics [width=0.36\columnwidth]{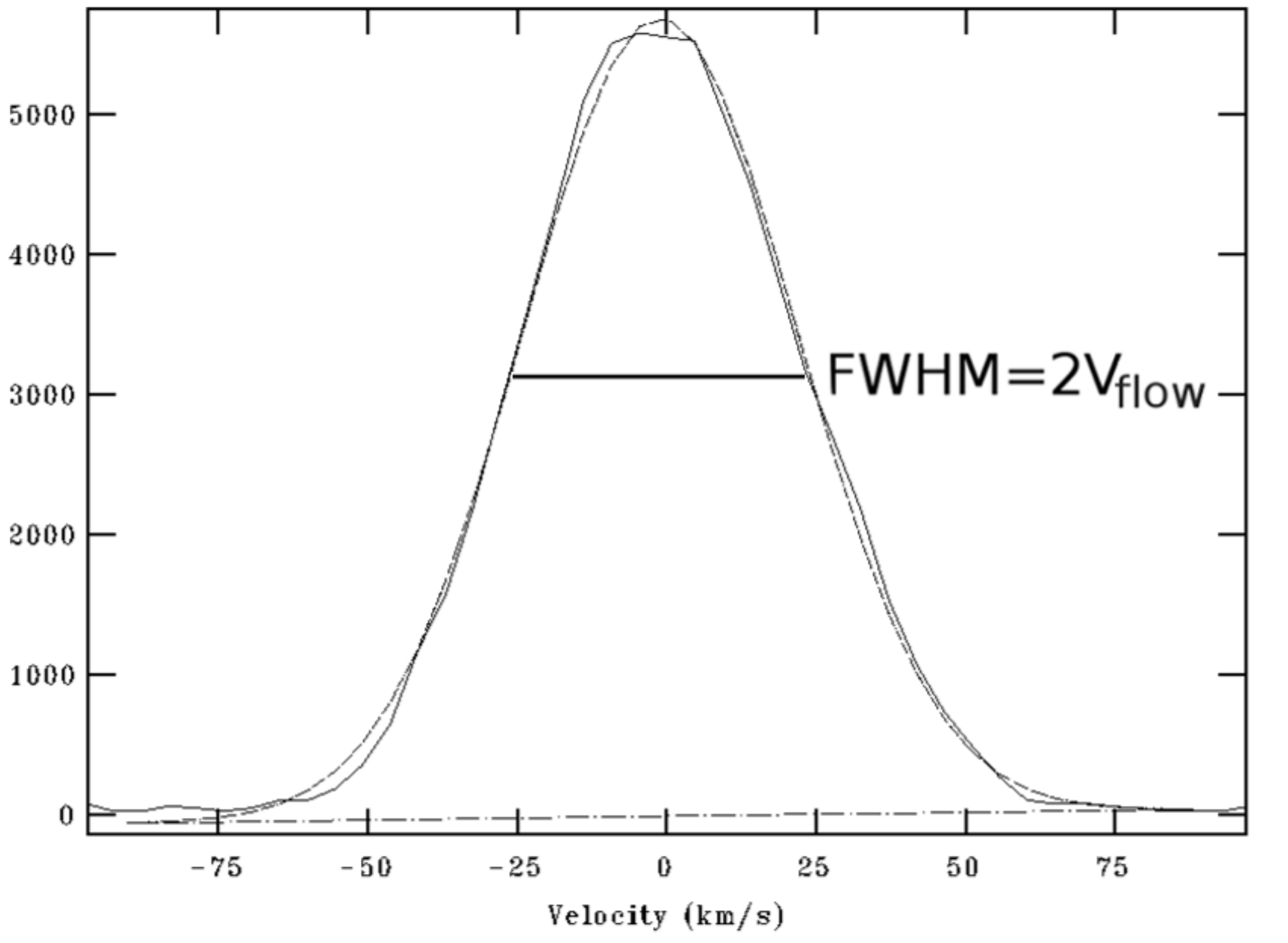} \\
\includegraphics [width=0.12\columnwidth]{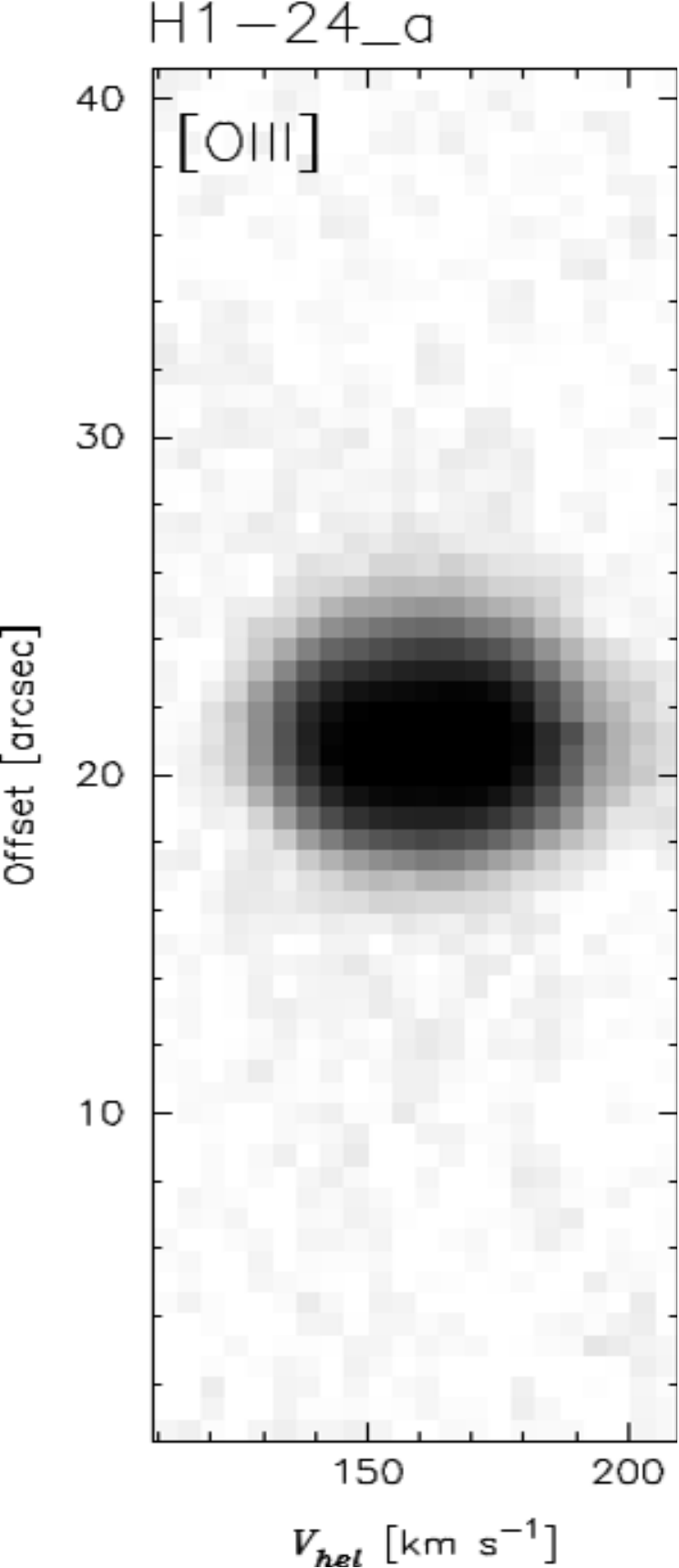}
\includegraphics [width=0.36\columnwidth]{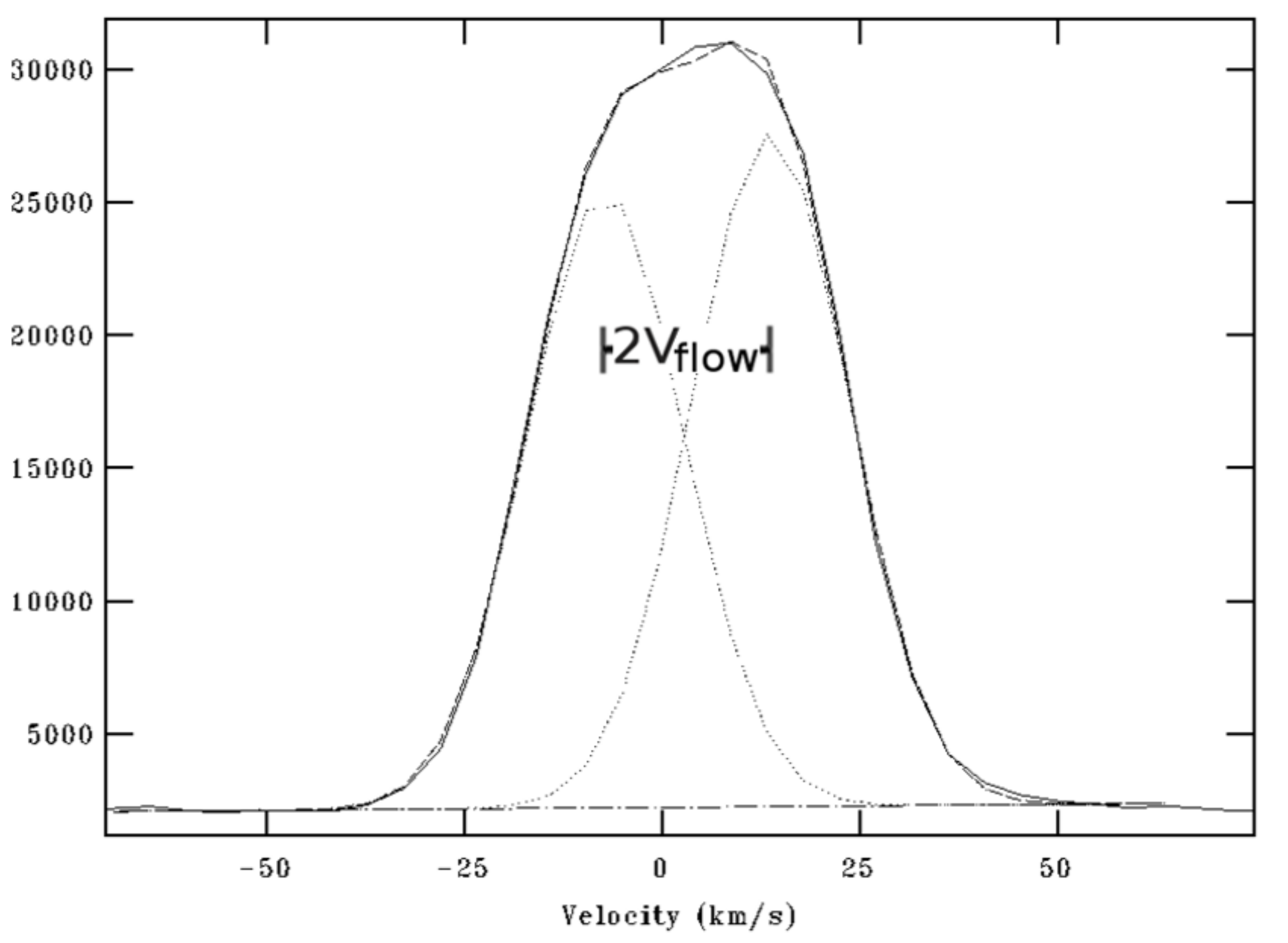} \\
\includegraphics [width=0.12\columnwidth]{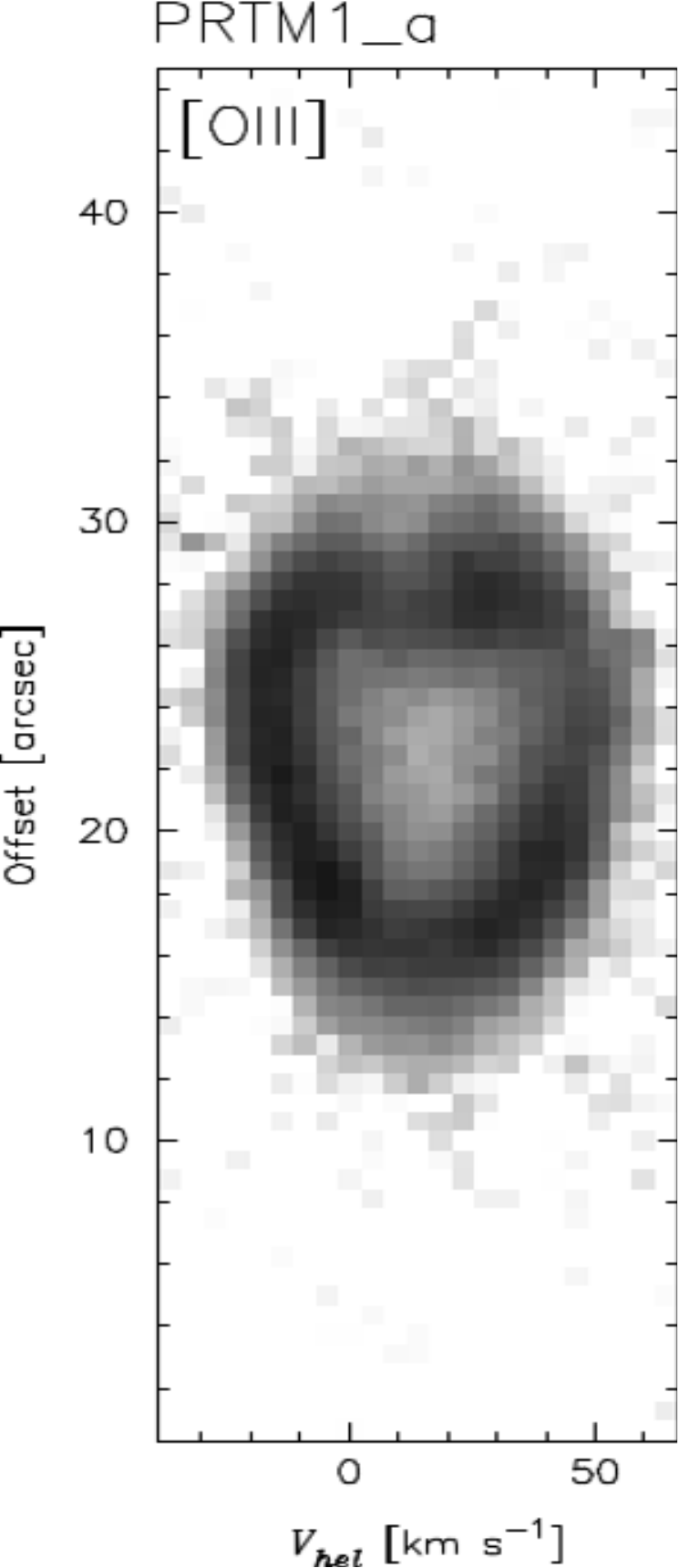}
\includegraphics [width=0.36\columnwidth]{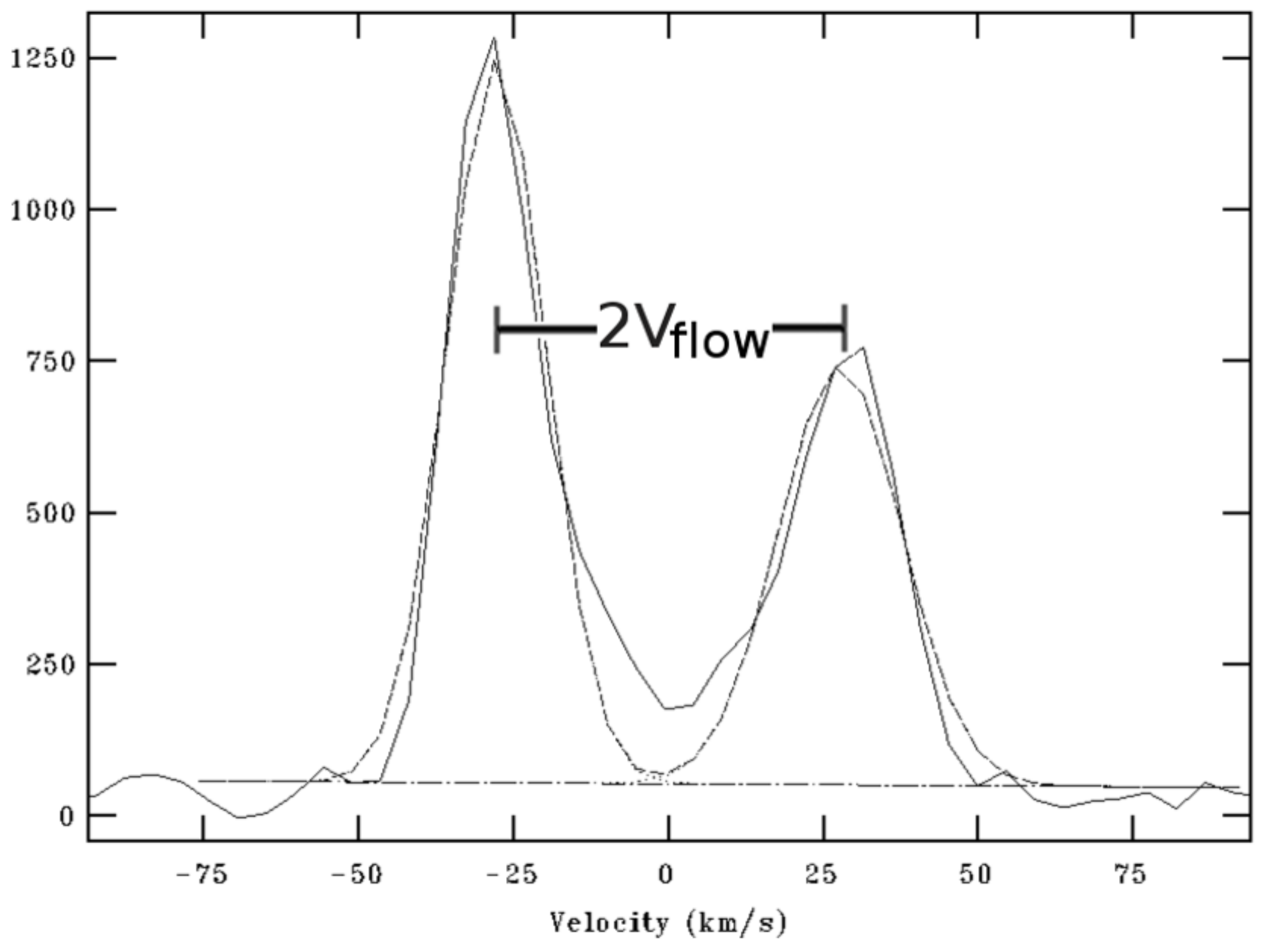}
%
%
\caption{ \small Data for PRMG 1, H 1-24 and PRTM 1 are shown to illustrate the different methods for determining the expansion velocities. Top panel: The [\ion{O}{3}] $\lambda$5007 line profile for PRMG 1 is filled and symmetric, so we fit a single Gaussian profile and assign half of the resulting FWHM as the expansion velocity. Middle and bottom panels: the H$\alpha$ line profiles for H 1-24 and PRTM 1 are marginally and clearly split, respectively, so we fit individual Gaussians to each component and the expansion velocity is taken as half the peak-to-peak difference in velocity between the blue- and red-shifted components at the point of maximum splitting. Depending upon the properties of the line profile, we used the appropriate method in each case.}
   \label{figure01}
\end{center}
\end{figure}

For the objects that are spatially unresolved by the spectrograph slit, the line profiles  fall into two categories. In the first, the line profile is a single symmetric profile. In the second case, additional or split components are apparent. When the profile is symmetric, we fit a single Gaussian profile and assign half of the resulting full width at half maximum intensity (FWHM) as the expansion velocity \citep[see e.g.][]{G&Z00}. In these cases, we correct the measured FWHM for instrumental broadening (11.5\,km\,s$^{-1}$ FWHM , thermal Doppler broadening (21.4\,km\,s$^{-1}$  and 5.3\,km\,s$^{-1}$ FWHM for H$\alpha$ and [\ion{O}{3}]), assuming a temperature of $10^4$\,K, and fine structure broadening in the case of the H$\alpha$ emission line \citep[7.5\,km\,s$^{-1}$ FWHM;][]{GD08} as in \citet{Richer08}. In Table \ref{table02}, we indicate the measurements obtained using this method with an asterisk.  For spatially-unresolved objects with asymmetric profiles or for spatially resolved objects with split profiles, we fit blue- and redshifted Gaussian components and assign $\mathrm V_{exp}$, as half the peak to peak difference in velocity between the blue- and redshifted components at the point of maximum line splitting, usually the centre of the line profile. Examples of these measurements are shown in Figure \ref{figure01}. We measured $\mathrm V_{exp}$ from all available emission lines for all the objects in the sample.

\section{RESULTS}\label{sec_sample_results}

\subsection{Velocity Distributions}

The velocity distributions from all the emission lines measured for both groups, halo and low metal PNe are shown in Figure 2. It is found that most of the PNe tend to concentrate at $V_{exp}$ $\lesssim$ 20\,km\,s$^{-1}$ in the two groups, independently of the emission line considered. In Table \ref{table03}, we present the average expansion velocity measured for each line of the two groups of PNe.  Since the number of emission lines observed for each PN varies, the number of objects considered to obtain the average expansion velocity will differ for each emission line (H$\alpha$, [\ion{N}{2}], and [\ion{O}{3}]).  Therefore, we obtained the average expansion velocities for the halo and low metal objects considering each emission line separately.  The average values and the number of objects used to calculate them, are shown in Table \ref{table03}.  No objects were excluded from these averages.  It is clear that, from all emission lines considered, the average values obtained for halo and low metal PNe are in excellent agreement with each other, clearly indicating that these groups share the same kinematic behavior or expansion velocity distribution.

The left panel of Figure \ref{figure03} presents the cumulative distribution functions (CDFs) for the expansion velocities of the halo and low metal PNe in the three emission lines.  For all three emission lines, the CDFs are similar.  Formally, a U-test \citep{walljenkins2003} indicates that the differences are not statistically significant and that the halo and low metal PNe arise from the same parent population with a probability of at least 38\%.  Since the kinematics or expansion velocity $\mathrm V_{exp}$ for the two groups are very similar, we henceforth consider the halo and low metal PNe as a single group. Hereafter, we will refer to this group simply as the sample.  The distribution of expansion velocities for the present sample is biased to low values, with 62\%, 70\%, and 87\% having expansion velocities below 20\,km\,s$^{-1}$ in the H$\alpha$, [\ion{O}{3}] $\lambda$5007 and [\ion{N}{2}] $\lambda$6584 lines, respectively.  This result indicates that the bulk of matter of the nebular shells of most of these PNe is indeed expanding at a relatively slow rate.

\subsection{Location in the H-R Diagram} \label{sec_H-R_Diagram}

Considering the selection process for the objects in our sample, we expect that the CSs  should be found along the evolutionary tracks corresponding to the lowest post-AGB masses in the H - R diagram. We adopted luminosities (L/L$_\odot$) and effective temperatures (T$_{eff}$) available from different sources  for individual PNe in our sample \cite[e.g., ][]{Kaler90, Pena92, Frew08, Otsuka10, STS10} to locate the CSs in our sample in the H-R diagram. 
We also include the luminosities and temperatures for H 1-24 and M 3-21 from \citet{zhangkwok1993}, which are model-dependent, but nominally independent of  distance.  These data for the CSs in this sample are heterogeneous.  For A 18, K 3-27, and BoBn 1, these parameters are based upon nebular properties, and only limits are available for A 18 and K 3-27.  For the other seven objects apart from H 1-24 and M 3-21, the temperatures and luminosities are based upon observations of the central stars and so may be more reliable.  We have compared the distances used above with the recent compilation from \citet{frewetal2016} and find that, for the halo PNe the distances from \citet{frewetal2016} agree to within $\pm 10$\%, for the low metallicity PNe the distances from \citet{frewetal2016} are typically larger by 30-37\%, except for PRMG 1, where the two distances differ by a factor of 3.  Given that the changes implied by these distances do not affect our results, we have kept our original distances.  The results are presented in Figure \ref{figure05}, where we also include the evolutionary tracks from \citet{VW94}. In spite of the usual uncertainties in deriving the CS parameters, the location of most of these CSs in the H - R diagram confirms that they indeed originate from very low mass progenitors.  

Figure \ref{figure06} emphasizes this result.  In Figure \ref{figure05}, we indicate the H$\alpha$ expansion velocities for the objects shown, save for GJJC 1 and PRTM 1 for which we use the [\ion{O}{3}] expansion velocities since they are the only spectra available for these objects.  In Figure \ref{figure06}, we also include the results from \citet{Pereyra13} for a sample of evolved PNe with more massive CSs \citep[whose stellar parameters all come from][]{Frew08}.  \citet{Pereyra13} divided their evolved PNe into two evolutionary classes, Mature and HE (highly evolved), terminology that we use in what follows.  Their Mature and HE PNe have high temperatures, but high and low luminosities, respectively, making the HE PNe the more evolved of the two classes.  The objects from the present sample are distributed near or below the 0.569 M$_\sun$ post-AGB track, whereas those from \citet{Pereyra13} are above this evolutionary track and extend towards higher masses.   It is clear in this figure how  the two samples dominate different regions in the diagram and also how they segregate in expansion velocity, with the objects from the present sample clearly showing the slowest expansion velocities.   The objects with higher velocities in the present sample are located at higher effective temperature and high luminosity, in approximately the same region occupied by the objects with the highest expansion velocities from \citet[Mature PNe;][]{Pereyra13} and \citet{Richer08, Richer10}.  However, the average expansion velocities for the objects with the highest values in each sample are very different. 
The average velocity for the objects from the \citet{Pereyra13} sample is $37$\,km\,s$^{-1}$ while, in the case of the present sample, the average velocity is $25$\,km\,s$^{-1}$ for the objects in the same region of the H-R diagram, which is a very substantial difference.

We may quantify the foregoing considering the CDFs for the objects within the rectangle in Figure \ref{figure06}.  
A U-test indicates that there is only a 3.5\% probability that the two distributions arise from the same parent distribution.  While not an overwhelmingly small probability, it is a small probability and supports the claim that the two populations are different.  Were we to extend the rectangle to include TS1, the probability would drop to 2.5\%.  


Regarding Figures  \ref{figure05} and \ref{figure06}, what is most relevant is the separation of the PNe from this study and from those from \citet{Pereyra13}.  Clearly, the masses associated with the majority of the objects from this study are lower than those for the PNe from \citet{Pereyra13}.  Undoubtedly, this is the result of the different selection criteria adopted by the two studies, and validates our selection criteria.  That the kinematics differ between these two samples is one of the key results of this study.  The masses associated with many of the PNe in our sample appear to be below the threshold generally thought to produce observable PNe, of $\sim 1$\,M$_{\odot}$ given a limiting CS mass of $\sim 0.55$\,M$_{\odot}$ to produce an observable PN \citep[e.g.,][]{schonberner83} and recent estimates of the initial-final mass relation \citep[e.g.,][and references therein]{kaliraietal09, dobbieetal12}.  However, the agreement, or lack thereof, between the masses derived from observational data and theoretical evolutionary tracks are not the result of our measurements or our selection criteria.  This mass mismatch deserves attention, but is not the subject of this paper.  The mass mismatch is not resolved if we use the masses from \citet{frewetal2016}.

\section{DISCUSSION}\label{sec_discussion} 

\subsection{The Kinematics of PNe from Low Mass Progenitors}

Previously, \citet{Richer08, Richer10} and \citet{Pereyra13} have found that the highest expansion velocities observed in PNe occur at high luminosity and high temperature.  Figure \ref{figure05} indicates that this trend is also true for the PNe from the present sample, chosen to select PNe that arise from stellar progenitors of the lowest masses.  Hence, it appears that the general kinematic evolution predicted by models (see \S 1) applies to the PNe derived from low-mass progenitors.  However, Figure \ref{figure06} indicates that the mass of the progenitor star does have a significant effect upon the expansion velocities observed, since the expansion velocities observed in the present sample are much lower than those observed for the sample of Mature PNe from \citet{Pereyra13}.  It has long been known that the wind velocities from lower mass OH/IR stars are lower than those of their higher mass counterparts \citep{Lewisetal1990}.  Likewise, both theory and observation indicate that the wind velocities from low-metallicity AGB stars are lower than their counterparts at higher metallicity \citep{Wood92, Marshall04, Mattsson08, Wachter08, Groenewegen09, Lagadec10}, though the mass-loss rates may not differ much.  Finally, the lower mass CSs in the current sample clearly have lower luminosities and wind energies than higher mass CSs \citep[e.g.,][]{Blocker95, Villaver02}. 
So, it is not surprising that we find substantially lower expansion velocities than have been found for other samples of PNe with higher mass CSs (see Figure \ref{figure06}).  

We quantify the foregoing by comparing the present sample with that studied by \citet{Pereyra13}.  The right panel of Figure \ref{figure03} presents the CDFs for the sample in each emission line and compares them with the CDFs for the two populations of evolved PNe studied by \citet{Pereyra13}.  We clearly see that the CDFs for the present sample are very different from those for the PNe from \citet{Pereyra13}.  Comparing the CDFs for the present sample and the highly evolved PNe from \citet{Pereyra13}, the probability that they arise from the same parent population is only $3.3\times 10^{-6}$, based upon the H$\alpha$ line.  The probability is even lower if we consider the mature PNe from their sample.  The distribution of expansion velocities for the present sample is shifted to lower values, with 62\%, 70\%, and 87\% having expansion velocities below 20\,km\,s$^{-1}$ in the H$\alpha$, [\ion{O}{3}] $\lambda$5007 and [\ion{N}{2}] $\lambda$6584 lines, respectively.  In the case of the highly evolved PNe \citep{Pereyra13}, only $17-24\%$ have such low velocities in the H$\alpha$ and [\ion{N}{2}] $\lambda$6584 lines and \emph{none} of their mature PNe have expansion velocities so low.  Therefore, the expansion velocities for the present sample are statistically different, shifted to lower values, from the expansion velocities of the samples studied by \citet{Pereyra13}.

By selecting halo and low-metallicity PNe we have selected a group of CSs of the lowest masses.  The expansion velocities of their nebular shells are the slowest among PNe in general, but their expansion velocities are nonetheless correlated with effective temperature. These results are in agreement with theory, which predicts that the mechanical luminosity of the stellar wind and the ionizing photon power output from the CS of a PN will be proportional to the remnant stellar mass after the AGB stage and both will increase during the constant luminosity portion of the post-AGB evolutionary track \citep{Perinotto04, Schonberner05a, Schonberner05b}. 
Accordingly we conclude that the kinematics of the PNe in the present sample behave consistently with previous findings \citep[and references therein]{Pereyra13} and that the low velocities found for these nebulae are a consequence of a weak CS wind driving the kinematics of the nebular shell of planetary nebulae that derive from stellar progenitors of low mass and low metal content.

These results cannot be compared directly to the recent models from  \citet[]{Schonberner10}, which indicate that PNe with low metallicity will result in lower cooling efficiency of the ionized shell, leading to a higher electron temperature and, as a consequence, higher expansion velocity of the nebular shell in comparison with similar nebulae with solar or higher metal content.  Their results are not applicable to the current sample because their models use the same CS and AGB envelope at all metallicities, whereas our sample contains CSs (and conceivably nebular shells) of lower masses than studied in other samples, i.e., for our sample, both the stellar mass and metallicity differ from those in other studies (see \S 2.1  and Figure \ref{figure06}).  The correlation of stellar age/mass with metallicity in the Milky Way prevents us from compiling a low-metallicity sample of PNe with CS masses similar to those in samples of PNe with higher metallicity for comparison.  
The results of \citet{Schonberner10} are in agreement with the results found for the brightest extragalactic PNe \citep{Richer10a}, where stellar mass is plausibly approximately constant while metallicity varies.  
  
\subsection{Velocity Measurements}


PNe can have complex geometries and dynamics, such as localized collimated outflows and plumes of low ionization regions,  dense cometary knots and double shell structures.  The latter, when present, involve the action of an X-ray emitting  hot bubble that produces a dense rim and an outer shell, in addition to a more extended halo.  Although widely predicted by theory, the presence of X-ray emitting hot bubbles has been detected so far only in a few, mostly elliptical, PNe with FLIER-type collimated outflows,  indicating that the development of a hot bubble in the evolution of a PN is not ubiquitous, but rather closely related to a narrow mass range of PNe cores with specific wind characteristics in strength and rate of evolution that indeed favor the formation of the hot bubble and the subsequent rim/shell structure \citep{Kastner12, freemanetal14}.

Therefore, when measuring radial (expansion) velocities, knowledge of the geometric structure of the object and spatially resolved spectroscopy are necessary tools for a proper \emph{interpretation} of the measurements.  Detailed kinematic studies of complex, individual objects take into account the different velocity patterns present in the nebular structure \citep[e.g.][]{Lopez12a, garciadiaz12}. For PNe with a rim/shell structure, \citet{Schonberner10} have pointed out that this case involves two different velocity patterns, one for the rim and one for the shell, with the rim moving slower than the shell and the former containing most of the matter, as predicted by the interacting winds model \citep{Kwok82, kw85}, but sometimes overlooked in the interpretation of the global PN expansion velocity.  In these cases, the global expansion has a non-ballistic structure and the shell's post-shock velocity is the closest approximation to a measure of the expansion rate, although this post-shock velocity cannot be measured directly \citep{Schonberner05b}.  

In their study of the evolution of the expansion rate of PNe at advanced stages of development, \citet{Pereyra13} specifically selected PNe without rims or filamentary structures, i.e. mainly old nebulae composed of a single, closely spherical, smooth shell where the overall expansion velocities  can be adequately determined from the matter with the highest emission measure within the spectrograph slit.  Considering that the slowly evolving, low metallicity PNe with low-mass central stars, the focus of the present study, are often spatially unresolved in our ground-based spectroscopy and are expected to have weak winds that are unlikely to produce a hot bubble and consequently a rim/shell structure \citep{Schonberner10}, we analyze the spectroscopic data here in a similar way as in \citet{Pereyra13}.

As stated in the introduction, the term  used here as \lq\lq expansion velocity", $\mathrm V_{exp}$, should not be confused with the post-shock velocity of the matter immediately behind the outer shock front.  $\mathrm V_{exp}$ is the emission-weighted expansion velocity for the matter projected within the spectrograph slit.  By this definition, $\mathrm V_{exp}$ is expected to increase monotonically at least until the CS achieves its maximum temperature, with little sensitivity to the chemical composition of the nebular shell \cite{Schonberner10}.  As such, $\mathrm V_{exp}$ is an adequate parameter for characterizing the evolution of nebular shells or their kinetic energy content, since it is a characteristic velocity for much of the mass in the nebular shell.  On the other hand, $\mathrm V_{exp}$ may be unsuitable for determining their kinematic ages or expansion parallaxes/distances, for which the outer shock velocity or other pattern velocity is preferred \citep[e.g.,][]{Jacob13}.  No single velocity can completely describe the motion of all of the matter in the nebular shell and the different velocities used for different purposes are not equivalent.  Rather, the strategy to measure the velocity should be chosen appropriately for the purpose for which the velocities are needed.  

As an illustration of the foregoing, there are five low-metallicity PNe common to our sample and the samples studied by \citet{Schonberner14} and \citet{Otsuka03, Otsuka09, Otsuka10, Otsuka15}: BoBn 1, H 4-1, PRMG 1, K 648/Ps 1, and DdDm 1.  We compare the results in Table \ref{table04}.  From Subaru HDS spectra, \citet{Otsuka09, Otsuka10, Otsuka15} measure, as we do, the emission-weighted velocity within the slit, and obtain results very similar to ours, for each line and object, differing by less than 2\ km\,s$^{-1}$, similar to our measurement uncertainties (Table \ref{table02}).  From high-resolution spectra, \citet{Otsuka03} fit multiple Gaussian components of which we adopt that referring to the large majority of the flux.  Their result is again similar to ours.  Using high-resolution VLT FLAMES/ARGUS spectra, \citet{Schonberner14} fit four components to each of their ([\ion{O}{3}]) line profiles, assuming the existence of a rim and shell.  Their results differ substantially from the others, often differing by more than 10\ km\,s$^{-1}$, as a result of both their method and their interests.  

\section{CONCLUSIONS}\label{sec_conclusions}

We analyzed 26 Planetary Nebulae (PNe) to characterize the kinematics of the nebular shells of PNe that appear to derive from progenitor stars of the lowest masses.  This is the largest sample used to date for this purpose and is composed of PNe from the Galactic halo as well as PNe of low metallicity.
The kinematical behavior of the two groups under study is very similar for all of the emission lines considered. Between 62\% and 87\% of the PNe in both groups have expansion velocities of $20$\,km\,s$^{-1}$ or less for all of the emission lines considered, indicating that the bulk of the entire mass in these objects is indeed expanding slowly.  
Of the 26  PNe in the sample, 12 of them have published luminosities and temperatures for their CSs, allowing them to be plotted in the H-R diagram. Comparing their location in the H-R diagram with the evolutionary tracks from \citet{VW94} confirms that these halo and low-metallicity PNe appear to derive from low mass progenitors.  Finally, the objects with the highest expansion velocities in our sample are located at higher effective temperature, as has been found previously. However, their average expansion velocity, $25$\,km\,s$^{-1}$, is very different from what was found for the Mature PNe from \citet[][$37$\,km\,s$^{-1}$]{Pereyra13} that are expected to have near-solar metallicities and derive from more massive progenitors.

Generally, the observed kinematical behavior of these PNe apparently derived from low mass progenitor stars is in excellent agreement with theoretical predictions and previous observations.  The nebular shells of most of the PNe in this sample are indeed expanding slowly, as expected for the weak winds observed and predicted for low-mass AGB stars and their progeny.  Likewise, the higher expansion velocities observed at high temperature in this sample are also consistent with previous findings, both theoretical and observed.  
We suggest that the low expansion velocities found for these PNe are most likely a consequence of  weaker winds from both their low-mass and low-metallicity AGB progenitors and the low-mass CSs driving the kinematics of their nebular shells.

\acknowledgments
The authors gratefully acknowledge financial support through grants from CONACyT (82066, 178253) and UNAM-PAPIIT (IN116908, IN110011). Special thanks are due to Dr. Teresa Garc\'ia D\'iaz for her excellent work with data calibration for the PNe in our sample. \\ M. Pereyra is also grateful to the Direcci\'on General de Estudios de Posgrado of UNAM, Schlumberger Foundation Faculty for the Future Fellowship Program and CONACyT (through the assistant researcher grant from the SNI) for additional financial support.  The authors thank the anonymous referee for useful comments.


\begin{deluxetable}{llccccccccccccc}
\tabletypesize{\scriptsize}
\rotate
\tablecaption{The final sample:  Halo PNe and PNe with $\log$\,(O/H)$+12\lesssim8.0$.} 
\scriptsize
\tablewidth{0pt}
\tablehead{
\colhead{Object}&\colhead{PN G}&\colhead{V$_{\bf H\alpha}$$^{\dagger}$}&\colhead{V$_{\bf [NII]}$$^{\dagger}$}&\colhead{V$_{\bf [O III]}$$^{\dagger}$}&\colhead{$(\mathrm O/\mathrm H)^{\ddag}$}&\colhead{Ref$^{\ddag}$}&\colhead{Group$^\bigstar$}&\colhead{$I(4686)^{\spadesuit}$}&\colhead{Ref$^\spadesuit$}&\colhead{log T (K)}&\colhead{log L/Lo}&\colhead{Ref$^\clubsuit$}\\
}
\startdata
A 18	&	216.0-00.2	&	10~~	&	16~~	&		&	7.99	& {\itshape e} &	LowMet	&	11.9	&	$^{3}$	&	5.03	&	1.86	&	{\itshape I}	\\
BD +332642 &	052.7+50.7	&	11 *	&	11 *	&		&		& {\itshape } &	Halo	&		&		&		&		&		\\
BoBn 1	&	108.4-76.1	&	23 *	&	19~~	&		&	7.89	& {\itshape f} &	Halo	&	20	&	$^{1}$	&	5.09	&	3.07	&	{\itshape II}	\\
DdDm 1	&	061.9+41.3	&	18 *	&	18 *	&	11~~	&	8.13	& {\itshape f} &	Halo	&	$\leq$2	&	$^{1}$	&	4.60	&	3.50	&	{\itshape III}	\\
GJJC 1	&	009.8-07.5	&		&		&	13 *	&		& {\itshape } 	&	Halo	&		&		&	4.69	&	3.34	&	{\itshape III}	\\
H 1-24	&	004.6+06.0 	&	16 *	&	12~~	&	18 *	&	7.52	& {\itshape b} &	LowMet	&	$\leq$4	&	$^{1}$	&	4.61	&	3.78	&	{\itshape VI}	\\
H 2-48	&	011.3-09.4	&	12 *	&	15 *	&	~7 *	&	7.89	& {\itshape g} &	LowMet	&	$\leq$1	&	$^{1}$	&		&		&		\\
H 4-1	&	049.3+88.1	&	13 *	&	13 *	&		&	7.85	& {\itshape a} &	Halo	&	10	&	$^{1}$	&		&		&		\\
JAFU 1	&	002.1+01.7	&	15~~	&	18~~	&	16~~	&		& {\itshape } &		Halo	&		&		&		&		&		\\
JAFU 2	&	353.5-05.0	&	28~~	&		&	25~~	&		& {\itshape } &		Halo	&		&		&		&		&		\\
K 3-27	&	061.0+08.0	&	23~~	&		&		&	7.96	& {\itshape h} &	LowMet	&	97	&	$^{1}$	&	4.98	&	3.20	&	{\itshape I}	\\
K 5-1	&	000.4+04.4	&	16 * 	&	12~~	&	18 *	&	7.88	& {\itshape d} &	LowMet	&		&		&		&		&		\\
K 5-5	&	001.5+03.6	&	17 *	&	20~~	&	16 *	&	7.72	& {\itshape d} &	LowMet	&		&		&		&		&		\\
M 3-21	&	355.1-06.9	&	14 *	&	13~~	&	13 *	&	8.00	& {\itshape g} &	LowMet	&	7.5	&	$^{1}$	&	4.65	&	3.80	&	{\itshape VI}	\\
NGC 4361&	294.1+43.6 	&	23~~	&		&		&	8.15	& {\itshape c} &	LowMet	&	115	&	$^{1}$	&	5.10	&	3.53	&	{\itshape IV}	\\
PRMG 1	&	006.0-41.9	&	26 *	&		&	26 *	&	8.06	& {\itshape c} &	LowMet	&		&			&	4.84	&	3.02	&	{\itshape III}	\\
PRTM 1	&	243.8-37.1	&		&		&	29~~	&	8.10	& {\itshape c} &	LowMet	&	$\leq$15&	$^{1}$	&	4.90	&	3.37	&	{\itshape III}	\\
PS 1 (K 648) &	065.0-27.3	&	17 *	&	12 *	&	17 *	&	7.85	& {\itshape f} &	Halo	&	$\leq$1	&	$^{1}$	&	4.54	&	3.48	&	{\itshape III}	\\
SaSt 2-3 &	232.0+05.7	&	10 *	&	10 *	&		&	8.29	& {\itshape i} &	Halo	&	$\leq$1	&	$^{1}$	&		&		&		\\
Sb 20	&	014.8-08.4	&	19~~	&		&	17~~	&	7.88	& {\itshape d} &	LowMet	&	30.2	&	$^{2}$	&		&		&		\\
Sb 32	&	349.7-09.1 	&	31~~	&	36~~	&	38~~	&	7.80	& {\itshape g} &	Halo	&	25.7	&	$^{2}$	&		&		&		\\
Sb 33	&	351.2-06.3	&	27~~	&	32~~	&		&	7.62	& {\itshape d} &	LowMet	&	17.5	&	$^{2}$	&		&		&		\\
SB 38	&	352.7-08.4	&	21~~	&		&	25~~	&	7.91	& {\itshape g} &	LowMet	&	65.6	&	$^{2}$	&		&		&		\\
Sb 42	&	355.3-07.5	&	17 *	&		&	16 *	&	7.82	& {\itshape d} &	LowMet	&	4.9	&	$^{2}$	&		&		&		\\
SB 55	&	359.4-08.5	&	20~~	&		&	20~~	&	7.58	& {\itshape g} &	LowMet	&	126.7	&	$^{2}$	&		&		&		\\
TS 01	&	135.9+55.9	&	22~~	&		&		&	6.82	& {\itshape k} &	Halo	&	77	&	$^{4}$	&	4.76	&	3.14	&	{\itshape V}	\\

\hline 
\enddata
 \scriptsize
	  \tablenotetext{*}{$\mathrm V_{exp}$ measurement obtained from the FWHM of a Gaussian fit.}
	  \tablenotetext{\dagger}{The uncertainty in these measurements is $\pm2$ km\,s$^{-1}$.}
	  \tablenotetext{\ddag}{This is the oxygen abundance on the scale $12+\log(\mathrm O/\mathrm H)$.  References:  a.-\citet{Costa96}, b.-\citet{Ratag97}, c.-\citet{Howard97}, d.-\citet{Escudero01}, e.-\citet{Costa04}, f.-\citet{Henry04}, g.-\citet{Gorny04}, h.-\citet{STG10}, i.- \citet{Pereira07}, j.-\citet{Henry10},  k.-\citet{STS10}. }
	  \tablenotetext{\bigstar} {The Halo and LowMet classifications correspond to Halo and Low Metallicity objects, respectively.}
	  \tablenotetext{\spadesuit}{This the He\,{\sc ii} $\lambda$4686 line intensity on a scale $I(\mathrm H\beta)=100$.  References: 1.-\citet{Tylenda94}, 2.-\citet{Escudero01}, 3.-\citet{Costa04}, 4.-\citet{STS10}.}
	  \tablenotetext{\clubsuit}{References for stellar parameters: {\itshape I}.-\citet{Kaler90}, {\itshape II}.-\citet{Otsuka10}, {\itshape III}.-\citet{Pena92}, {\itshape IV}.-\citet{Frew08}, {\itshape V}.-\citet{STS10}, {\itshape VI}.-\citet{zhangkwok1993}.} 
\label{table02}	  
\end{deluxetable}

\begin{table*}[]
\centering 
\caption{Average expansion velocities for Halo and Low Metal PNe.} 
\scriptsize
\begin{tabular}{c| c c| c c| c c}
\hline \hline
       &                          &             &                          &             &                           &             \\
Group  & $\overline{V}_{\bf[N II]} $$^\dag$ & No. Objects & $\overline{V}_{\bf H\alpha}$$^\dag$ & No. Objects &  $\overline{V}_{\bf [O III]}$$^\dag$ & No. Objects  \\
       &   & considered &   & considered &   & considered \\
\hline 
Halo PNe & 17  & 8 & 19 & 10 & 20 & 6 \\
Low Metal PNe  & 17 & 7 & 19  & 14  & 18 & 11 \\
\hline 
\end{tabular}
\tablenotetext{\dag}{\ The uncertainty in the measurements in $\pm2$ km\,s$^{-1}$.}
 \scriptsize
\label{table03} 
\end{table*}

\begin{table*}[]
\centering 
\caption{Comparison of recent kinematic observations of halo PNe.} 
\scriptsize
\begin{tabular}{l c c c c c c c}
\hline \hline
       & \multicolumn{3}{c}{Present work} & \citet{Schonberner14} & \multicolumn{3}{c}{Otsuka et al.} \\
       \cline{2-4} \cline{6-8}
Object  & H$\alpha$ & \boldmath[NII] & \boldmath[OIII] & \boldmath[OIII] & \boldmath HI & \boldmath[NII] & \boldmath[OIII] \\
       & (km\,s$^{-1}$) & (km\,s$^{-1}$) & (km\,s$^{-1}$) & (rim/post-shock; km\,s$^{-1}$) & (km\,s$^{-1}$) & (km\,s$^{-1}$) & (km\,s$^{-1}$) \\
\hline 
BoBn 1 & 23 & 19 & & 12/52 & 23$^a$ & 19 & 21 \\
DdDm 1 & 18 & 18 & 11 & & 18$^b$ & 19 & 11 \\
H 4-1  & 13 & 13 & & & 15$^c$ & & 18 \\
K 648/Ps 1 & 17 & 12 & 17 & 8/35 & 15$^d$ & 14 & 16 \\
PRMG 1 & 26 & & 26 & 12/49 & & & \\
\hline 
\end{tabular}
\tablenotetext{a}{\ \citet{Otsuka10} measure H$\beta$.}
\tablenotetext{b}{\ \citet{Otsuka09} measure H$\alpha$.}
\tablenotetext{c}{\ \citet{Otsuka03} measure H$\alpha$.  The data are for a position angle of 90$^{\circ}$, as for our observations.}
\tablenotetext{d}{\ \citet{Otsuka15} measure 26 \boldmath HI Balmer lines.}
\label{table04} 
\end{table*}

\begin{figure*}[]
\begin{center}
%
%
\includegraphics [width=0.49\columnwidth]{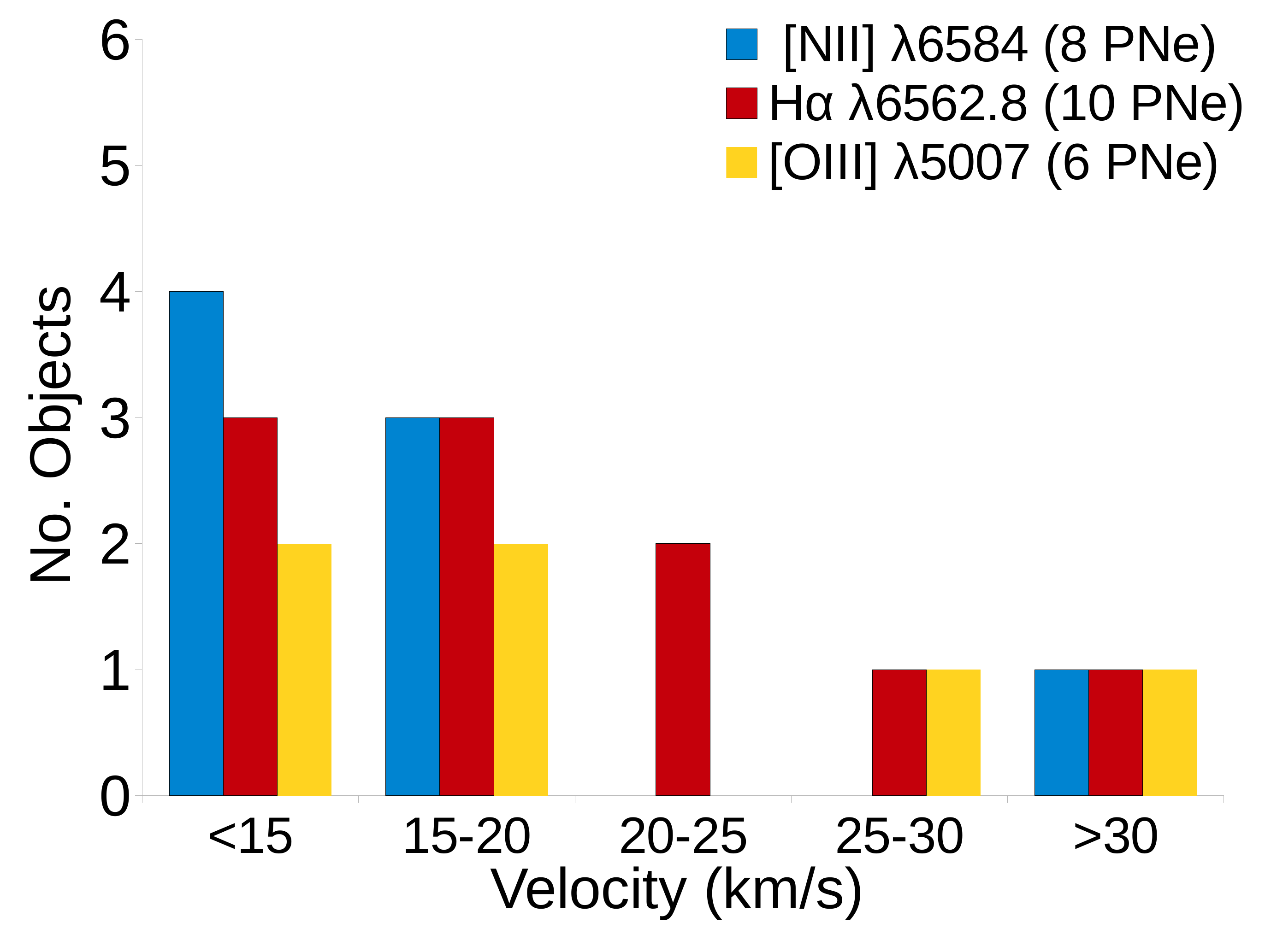} 
\includegraphics [width=0.49\columnwidth]{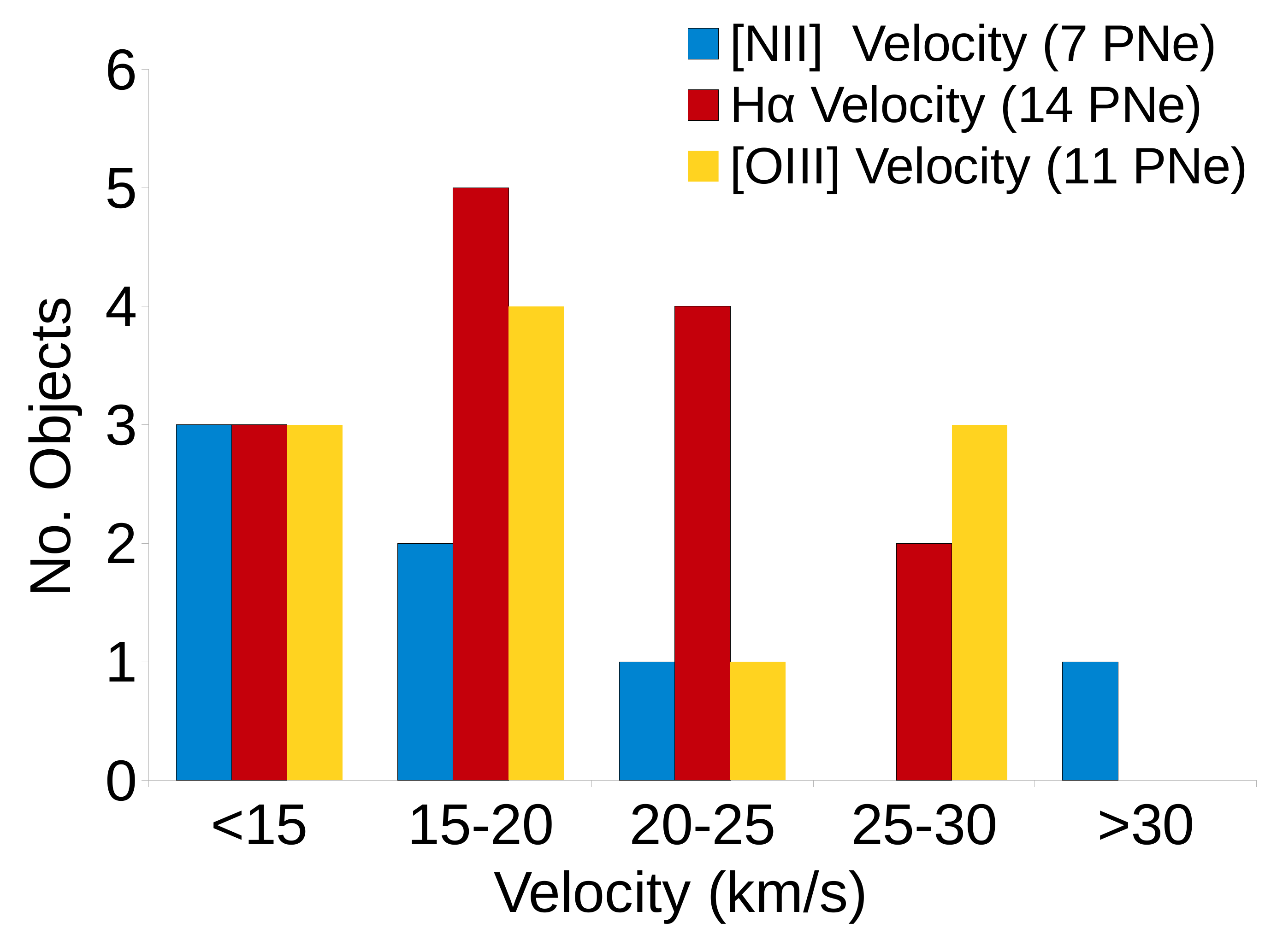} 
%
%
 \caption{\small We present the distribution of expansion velocities for the halo (left panel) and low metal (right panel) objects. For both groups, we plot the measurements obtained from each emission line (see symbol legend). The plots suggest that, no matter which emission line is considered, most of the objects have low expansion velocities.} 
   \label{figure02}
\end{center}
\end{figure*}

\begin{figure*}[]
\begin{center}
%
%
\includegraphics [width=0.49\columnwidth]{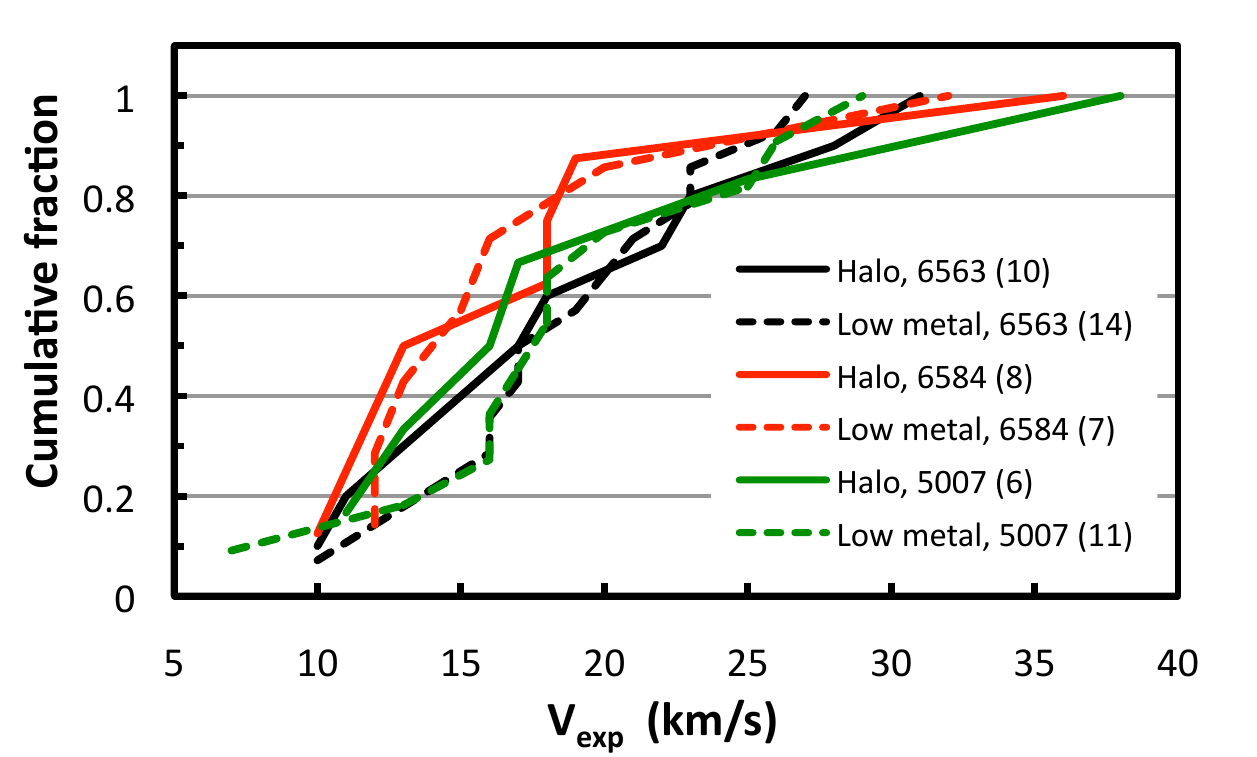} 
\includegraphics [width=0.49\columnwidth]{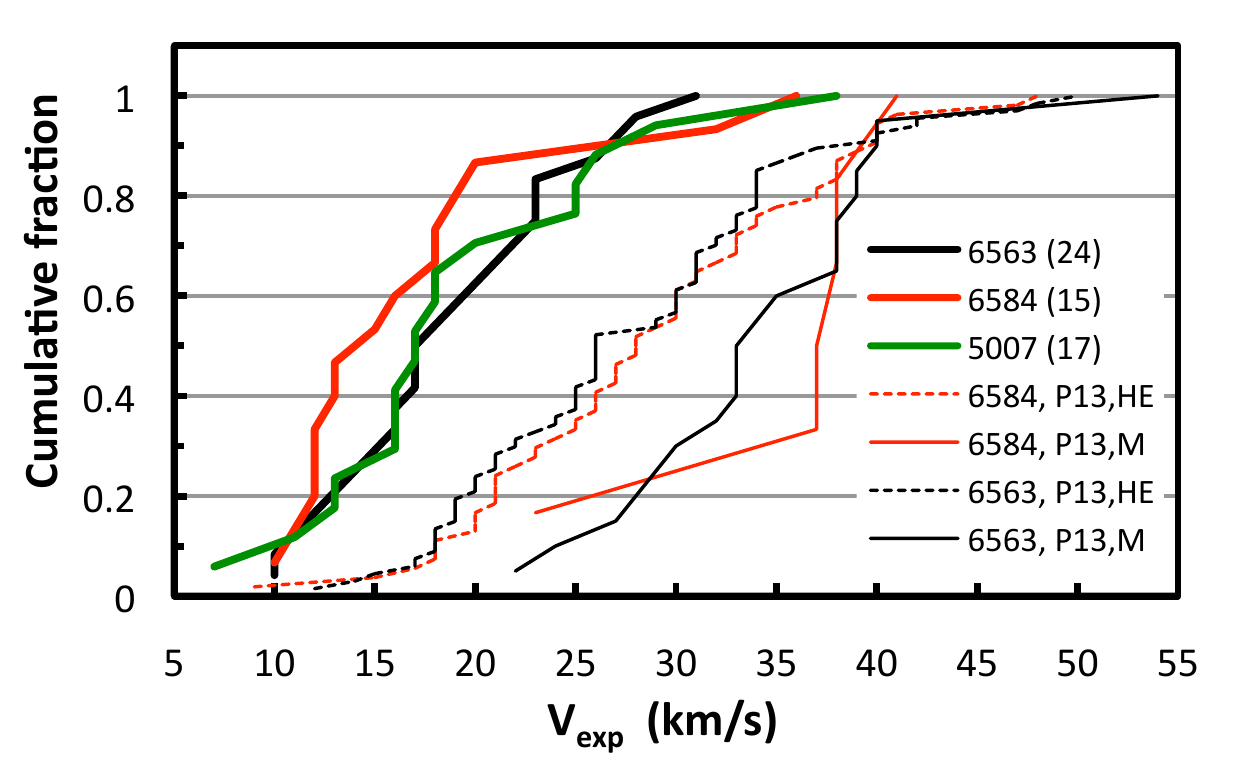}
%
%
 \caption{\small In the left panel, we present cumulative distribution functions for the expansion velocities in  H$\alpha$ $\lambda$6562.8 (black lines), [\ion{N}{2}] $\lambda$6584 (red lines), and [\ion{O}{3}] $\lambda$5007 (green lines) for the halo PNe (continuous lines) and low metal PNe (dashed lines).  The numbers in parentheses indicate the number of objects from the present sample.  It is clear that the kinematical behavior of the two groups is very similar in all emission lines.  For each emission line, a U-test \citep{walljenkins2003} indicates at least a 38\% probability that the halo and low metal PNe arise from the same parent population.  In the right panel, we present the cumulative distribution functions for our entire sample (thick lines) and the two populations studied by \citet[][thin continuous and dotted lines]{Pereyra13}.  The color scheme is the same as in the left panel.  A U-test indicates that the probability that the present sample and the highly evolved (HE) PNe from \citet{Pereyra13} arise from the same parent population is only $3.3\times 10^{-6}$, based upon the H$\alpha$ line.  The probability would be even lower for their mature (M) PNe.  In the present sample, 70-87\% of the PNe have expansion velocities below 25 km\,s$^{-1}$ depending upon the emission line considered.} 
   \label{figure03}
\end{center}
\end{figure*}

\begin{figure*}[]
\begin{center}
\includegraphics [width=5.5in]{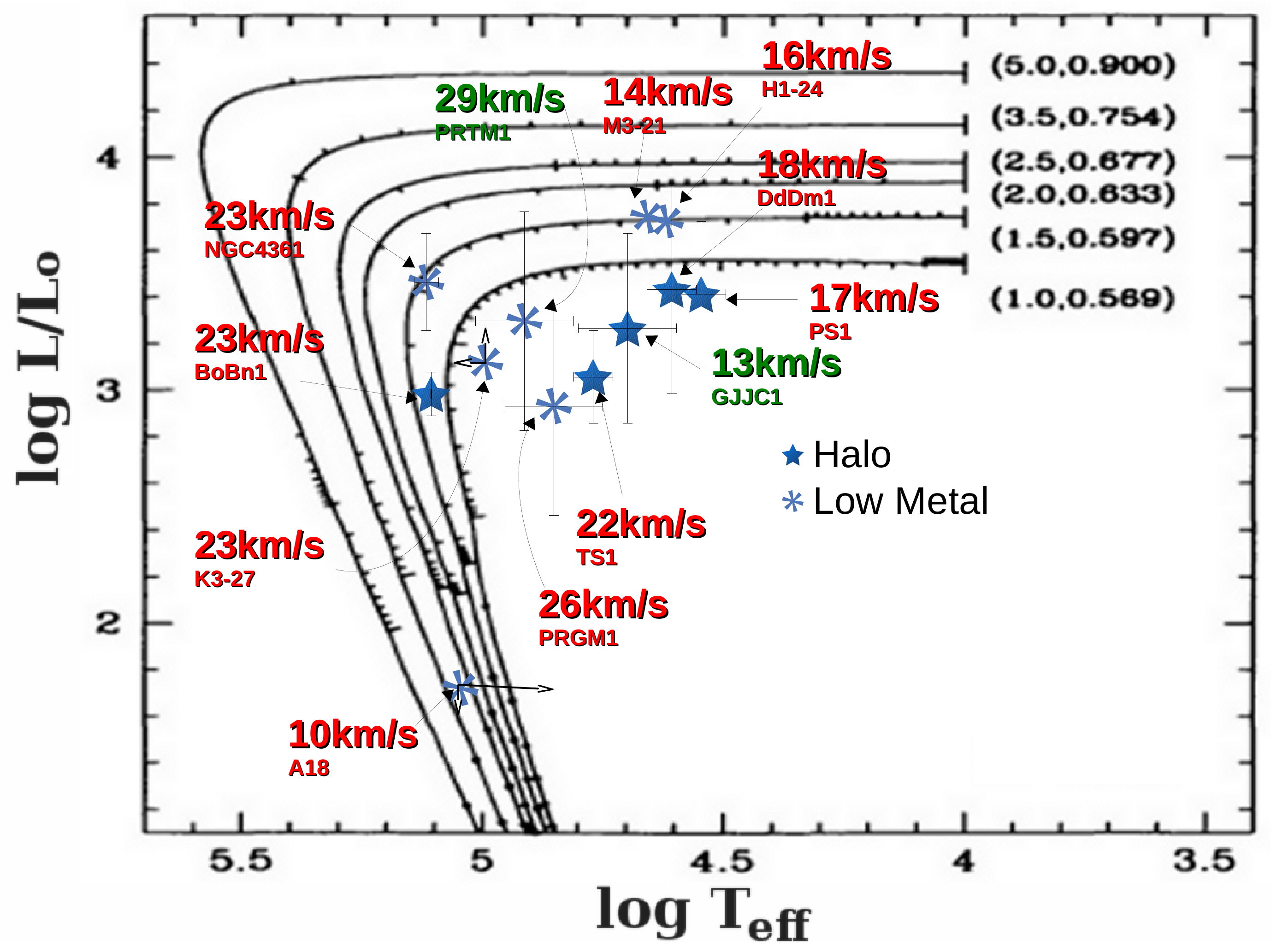}
\caption{\small We locate the sample PNe in the H-R diagram for objects where temperatures and luminosities are available for their CSs.  (For A18 and K 3-27, only limits are available.)  There is a clear concentration towards or below the  post-AGB track of lowest mass.  The evolutionary tracks are from \citet{VW94}.  The expansion velocities are for the H$\alpha$ line, except for GJJC1 and PRTM1, which are from the [\ion{O}{3}] $\lambda$5007 line.} 
   \label{figure05}
\end{center}
\end{figure*}

\begin{figure*}[]
\begin{center}
\includegraphics [width=5.5in]{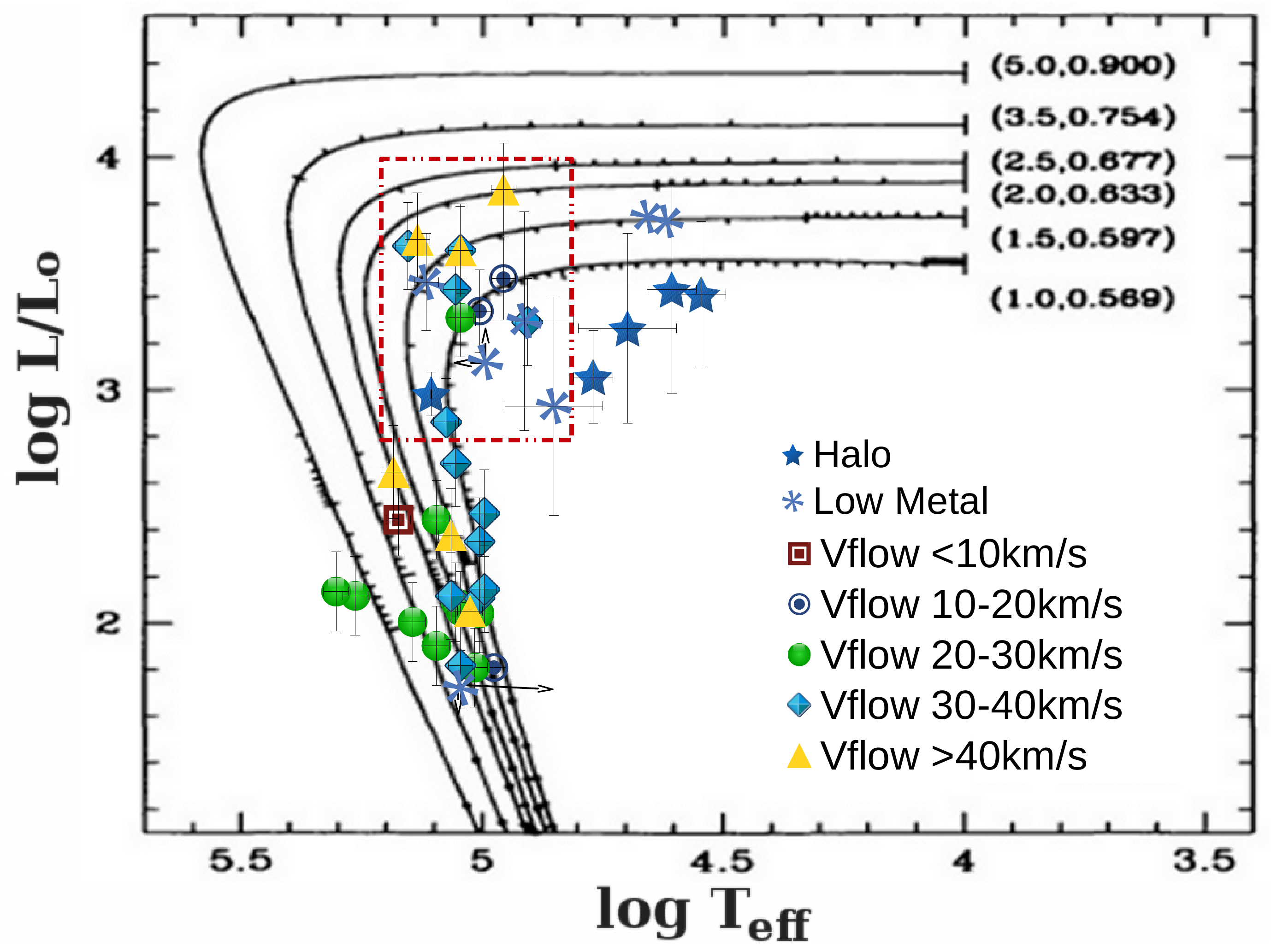}
\vspace*{-0.5cm}
\caption{\small We compare the objects from the present sample with those studied by \citet{Pereyra13} that should be derived from stellar progenitors of higher mass.  For the objects from this study, the expansion velocities are indicated in Figure \ref{figure05}.   For the objects from \citet{Pereyra13}, the expansion velocities are indicated by the symbols given in the legend.  Although there are few data, it is clear that the PNe in the present sample cluster towards lower masses than those from \citet{Pereyra13}.  The box includes the objects from both samples that have high effective temperature and high luminosity.  These have the highest expansion velocities in each sample, but the average values are very different.  We plot the H$\alpha$ measurements for all objects, except for GJJC 1 and PRTM 1, for which only [\ion{O}{3}] spectra are available.  } 
   \label{figure06}
\end{center}
\end{figure*}

\end{document}